\documentclass[12pt,oneside,final]{amsart}

\usepackage[utf8x]{inputenc}

\usepackage{mystyle}

\begin{document}

\title[Einstein-Hilbert Gravity Embedded in a Higher Derivative Model II]{On the Perturbative Quantization of Einstein-Hilbert Gravity Embedded in a Higher Derivative Model II }

\author{Steffen Pottel}
\address[S. Pottel]{22 Upper Ground, SE1 9PD London, United Kingdom}

\author{Klaus Sibold}
\address[K. Sibold]{Institute for Theoretical Physics, Leipzig University, Germany. \emph{Email address: }\rm \texttt{sibold(at)physik.uni-leipzig.de}.}

\date{\today}

\begin{abstract}
In a previous paper we presented the renormalization of Einstein-Hilbert gravity under inclusion of higher derivative terms and proposed a projection down to the physical state space of Einstein-Hilbert.
In the present paper we describe this procedure in more detail via decomposing the original double-pole field $h^{\mu\nu}$ in the bilinear field sector into a massless and a massive spin two field.
Those are associated with the poles at zero mass resp.\ at non-zero mass of $h$
in the tree approximation.
We show that the massive fields have no poles in higher orders hence do not 
correspond to particles. $S$-matrix unitarity is violated only in tree 
approximation.
On the way to these results we derive finiteness properties which are valid in 
the Landau gauge. Those simplify the renormalization group analysis of the model considerably. We also establish a rigid Weyl identity which represents a
proper substitute for a Callan-Symanzik equation in flat spacetime.
\end{abstract}

\maketitle

\tableofcontents

\section{Introduction}
Why would we add higher derivative terms (hds) to the Einstein-Hilbert action (EH) of gravity?
The answer goes back to the enormeously successful history of particle physics.
That started with Fermi's model of four-fermion-coupling, eventually in the form of V-A currents.
It was obvious that this was at most an effective theory because it violated tree unitarity of the $S$-matrix.
Hence one invented massive intermediate vector bosons.
Still, this was not yet power counting renormalizable, became so, however, via the Higgs effect.
Indeed, with this realization of masses for the intermediate bosons one could prove unitarity of the $S$-matrix: the electroweak standard model
was established.
The completion, even unification, with QCD yielded a theory which withstood all experimental tests as of yet.\\
In the context of gravity it was the fundamental insight by Einstein that the gravitational field, the metric $g^{\mu\nu}$, should couple to the energy-momentum-tensor (EMT) of matter.
In a perturbative version, where $g^{\mu\nu}=\eta^{\mu\nu}+h^{\mu\nu}$ with $h$ understood as ``small'' deviation from flat space it is $h^{\mu\nu}$ which couples to the matter EMT.
In a quantum field theoretic context ``smallness'' of a quantum field is, of 
course, not a meaningful notion, but is implemented here as perturbation theory 
which progresses in an expansion in the number of $h$ fields in addition to the
conventional one in number of loops.
The most important fact is, however, that the EMT has canonical dimension four.
If one aims at powercounting renormalizability as a minimal prerequisite for full renormalizability, which does not only mean ``removing infinities'', but rather satisfying standard axioms, then one must assign to $h$ canonical dimension zero.
But since renormalization brings in a rule to admit all terms compatible with canonical dimension less or equal to four in an action, one has to admit in the context of EH not only the standard term $\int\sqrt{-g}R$ but also $\int\sqrt{-g}R^{\mu\nu\rho\sigma}R_{\mu\nu\rho\sigma}$, $\int\sqrt{-g}R^{\mu\nu}R_{\mu\nu}$, and $\int\sqrt{-g}R^2$ \cite{Stelle}.
Together with the cosmological term they constitute a basis on the integrated 
level. \\
\noindent
For the construction of the model in perturbation theory, we applied the Bogoliubov-Parasiuk-Hepp-Zimmermann-Lowenstein (BPHZL) renormalization scheme 
\cite{pottel2021perturbative}.
The backbone of this scheme is power counting.
Important for a qualitative understanding is the role which the different classical invariants play in this context: in order to have off-shell IR-convergence the EH term must be present; in order to have UV-convergence the higher derivative terms are required.
This fact constrains tightly possible limits in terms of the couplings $c_3\kappa^{-2}$, $c_1$, and $c_2$.
The cosmological term is irrelevant in this respect. We suppress it by a 
normalization condition.

In the present paper we deepen the previous analysis by some more detailed study.
As a substitute of scaling in flat spacetime we derive a Ward identity for rigid
Weyl invariance. We show that in Landau gauge certain finiteness properties can be established which have been known in the context of Yang-Mills theories before \cite{Blasi:1990xz}.
This simplifies considerably the renormalization group (RG) analysis.
Central, however, is a discussion of the state space of the model.
By using spin two fields which decompose and diagonalize the sector which is 
bilinear in the $h$ field, \cite{stelle1978classical}, we can study the role, 
in particular the removal of those degrees of freedom which have been introduced
by the higher derivatives. We show that the massive fields have no poles in 
higher orders.
Potential respective singularities would be RG dependent, hence unphysical.
This implies that the violation of unitarity of the $S$-matrix is restricted to the tree approximation.

\section{Preliminaries}
In order to make the present paper reasonably self contained we 
collect the main formulae from paper I \cite{pottel2021perturbative}.
The starting point is the following action together with the variations of 
the fields.
\begin{align}\label{ivc}
\Gamma^{\rm class}&=\Gamma^{\rm class}_{\rm inv}+\Gamma_{\rm gf}+
                    \Gamma_{\phi\pi}+\Gamma_{\rm e.f.}\\
	\Gamma^{\rm class}_{\rm inv}&=\int\sqrt{-g}(c_3\kappa^{-2}R+c_2 R^2
                                           +c_1 R^{\mu\nu}R_{\mu\nu})\\
\Gamma_{\rm gf}&=-\frac{1}{2\kappa}\int g^{\mu\nu}(\partial_\mu b_\nu
                                       +\partial_\nu b_\mu)
	-\frac{1}{2}\alpha_0\int \eta^{\mu\nu}b_\mu b_\nu\\
\Gamma_{\phi\pi}&=-\frac{1}{2}\int(D^{\mu\nu}_\rho c^\rho)
	(\partial_\mu\bar{c}_\nu+\partial_\nu\bar{c}_\mu)\\
	D^{\mu\nu}_\rho&\equiv -g^{\mu\lambda}\delta^\nu_\rho\partial_\lambda	                              -g^{\nu\lambda}\delta^\mu_\rho\partial_\lambda
	                       +\partial_\rho g^{\mu\nu}\\
\Gamma_{\rm e.f.}&=\int(K_{\mu\nu}sh^{\mu\nu}+L_\rho sc^\rho)\\
	sg^{\mu\nu}&=\kappa D^{\mu\nu}_\rho c^\rho
	\quad sc^\rho=-\kappa c^\lambda\partial_\lambda c^\rho
	\quad s\bar{c}_\rho=b_\rho\quad  sb_\rho=0\\
	h^{\mu\nu}&=g^{\mu\nu}-\eta^{\mu\nu}\\
	 s_0h^{\mu\nu}&=-\kappa(\partial^\mu c^\nu+\partial^\nu c^\mu)\quad
	 s_1h^{\mu\nu}=-\kappa(\partial_\lambda c^\mu h^{\lambda\nu}
	                           +\partial_\lambda c^\nu h^{\mu\lambda}
				   -c^\lambda\partial_\lambda h^{\mu\nu})
\end{align}

Furthermore, we need the Slavnov-Taylor identity (ST). 
Since the s-variations of $h$ and $c$ are non-linear in the fields, 
they are best
implemented in higher orders via coupling to external fields, hence 
the ST identity then reads
\be\label{fbrst}
\mathcal{S}(\Gamma)\equiv
\int(\frac{\delta\Gamma}{\delta{K}}\frac{\delta\Gamma}{\delta h}
+\frac{\delta\Gamma}{\delta L}\frac{\delta\Gamma}{\delta c}
+b\frac{\delta\Gamma}{\delta\bar{c} })=0 .
\ee
The $b$-equation of motion
\be\label{beq}
\frac{\delta \Gamma}{\delta b^\rho}=
                     \kappa^{-1}\partial^\mu h_{\mu\rho}-\alpha_0b_\rho
\ee
is linear in the quantized field $b$ and can be integrated trivially to the 
original gauge fixing term. Thus it turns out to be useful to introduce a functional
$\bar{\Gamma}$ which does no longer depend on the $b$-field:
\be\label{Gmmbr}
\Gamma=\Gamma_{\mathrm{gf}}+\bar{\Gamma} .
\ee
One finds 
\be\label{rstc}
\kappa^{-1}\partial_\lambda\frac{\delta\bar{\Gamma}}{\delta K_{\mu\lambda}}
+\frac{\delta\bar{\Gamma}}{\delta\bar{c}_\mu} =0
\ee
as restriction. Hence $\bar{\Gamma}$ depends on $\bar{c}$ only via
\be\label{sceH}
H_{\mu\nu}=K_{\mu\nu} - \frac{1}{2\kappa}(\partial_\mu\bar{c}_\nu+\partial_\nu\bar{c}_\mu)
\ee
and the ST identity takes the form
\begin{eqnarray}\label{brGm}
\mathcal{S}(\Gamma)&=&\frac{1}{2}\mathcal{B}_{\bar{\Gamma}}\bar{\Gamma}=0\\
	\mathcal{B}_{\bar{\Gamma}}&\equiv& 
	\int(
  \frac{\delta\bar{\Gamma}}{\delta H}\frac{\delta}{\delta h}
+ \frac{\delta\bar{\Gamma}}{\delta h}\frac{\delta}{\delta H} 
+ \frac{\delta\bar{\Gamma}}{\delta L}\frac{\delta}{\delta c}
+ \frac{\delta\bar{\Gamma}}{\delta c}\frac{\delta}{\delta L}
	) .
\end{eqnarray}
This form shows that $\mathcal{B}_{\bar{\Gamma}}$ can be interpreted as a variation und thus
(\ref{brGm}) expresses an invariance for $\bar{\Gamma}$.

The free parameters of the model can be prescribed by the following normalization conditions on the vertex functions and their coefficients $\gamma^{(2)}_{\rm TT}$ and $\gamma^{(0)}_{\rm TT}$, resp.,
\begin{eqnarray}\label{highnorm}
\frac{\partial}{\partial p^2}\,\gamma^{(2)}_{\rm TT\,|{\substack{p=0 \\ s=1} }}&=
				   &c_3\kappa^{-2}\\
\frac{\partial}{\partial p^2}\frac{\partial}{\partial p^2}\,
	\gamma^{(2)}_{\rm TT\,|{\substack{p^2=-\mu^2\\ s=1} }}&=&-2c_1\\
\frac{\partial}{\partial p^2}\frac{\partial}{\partial p^2}\,
	\gamma^{(0)}_{\rm TT\,|{\substack{p^2=-\mu^2 \\ s=1} }}
                                        	&=&2(3c_2+c_1)\\
	\Gamma_{h^{\mu\nu}} &=&-\eta_{\mu\nu}c_0=0\\			
\frac{\partial}{\partial p_\sigma}
	\Gamma_{K^{\mu\nu}c_\rho|{\substack{p^2=-\mu^2 \\ s=1} }}&=&
                                -i\kappa(\eta^{\mu\sigma}\delta^\nu_\rho
	                          +\eta^{\nu\sigma}\delta^\mu_\rho
		  -\eta^{\mu\nu}   \delta^\sigma_\rho)\label{highnorm1} \\
\frac{\partial}{\partial p^\lambda}
	\Gamma_{{L_\rho}c^\sigma c^\tau|{\substack{p^2=-\mu^2 \\ s=1} }}&=&
			-i\kappa(\delta^\rho_\sigma\eta_{\lambda\tau}
				  -\delta^\rho_\tau\eta_{\lambda\sigma}).
\end{eqnarray}
The first four conditions fix the couplings, whereas the last two fix 
the field amplitudes of $h,K$ and $c,L$, resp.
Imposing the $b$-equation of motion (\ref{beq}) fixes $\alpha_0$ 
and the $b$-amplitude. \\ 

As a partial differential equation which is symmetric wrt.\ ST we 
derived the renormalization group equation (RG)
\be\label{rg1}
\mu\partial_\mu\Gamma_{|s=1}=
         (-\beta^{\rm RG}_{3} c_3 \partial_{{c_3}}
          -\beta^{\rm RG}_{c_1} c_1 \partial_{{c_1}}
          -\beta^{\rm RG}_{c_2} c_2 \partial_{{c_2}}
	  +\gamma^{\rm RG}_h\,\mathcal{N}_H
	  +\gamma^{\rm RG}_c\,\mathcal{N}_L)\Gamma_{|s=1} \,.
 \ee
 Here the $\mathcal{N}$'s are symmetric generalized leg counting operators
\be\label{lgct}
\mathcal{N}_H\equiv N_h-N_K-N_b-N_{\bar{c}}+2\alpha_0\partial_{\alpha_0}
                                           +2\chi\partial_\chi
\qquad \mathcal{N}_L\equiv N_c-N_L \,.
\ee

\section{Part I: Theory formulated in terms of $h$}

\subsection{Rigid Weyl Invariance}
Here we address the issue of replacing scaling in ordinary flat spacetime by a version appropriate for curved spacetime.
In flat spacetime scaling can be realized by the Callan-Symanzik equation (CS).
In \cite[eq.\ (309)]{pottel2021perturbative} we proposed a candidate for that purpose 
\begin{multline}\label{csqt}
(\mu\partial_\mu-\kappa\partial_\kappa-2c_3\partial_{c_{3}}
-(N_b-2\alpha_0\partial_{\alpha{_0}})+N_K+N_L
+\beta^{\rm CS}_1c_1\partial_{c_{1}}
+\beta^{\rm CS}_2c_2\partial_{c_{2}}
\\-\gamma^{\rm CS}_h\mathcal{N}_H
-\gamma^{\rm CS}_c\mathcal{N}_L)\Gamma_{|s=1}=
	\alpha^{\rm CS}\lbrack\kappa^{-2}\int\sqrt{-g}R\rbrack^3_3\cdot\Gamma_{|s=1} \,.
\end{multline}
This equation is correct, but since its rhs starts with order O($\hbar^1$), i.e.\
has no contribution in the tree approximation, it does not correspond to 
a CS equation in the standard sense.
Now, scaling of coordinates is anyway not a proper concept in the context of general relativity, it rather has to be replaced by Weyl transformations of the metric tensor 
$\delta^{\rm W}g^{\mu\nu}=2\sigma g^{\mu\nu}$ ($\sigma=\,\mathrm{const}$ is coined 
 ``rigid'').
We therefor study the identity
\begin{multline}\label{rgWl}
	\frac{1}{2\sigma}W^{\rm W}_{\rm rig}\Gamma\equiv \int(
g^{\mu\nu}\frac{\delta}{\delta g^{\mu\nu}}
-K_{\mu\nu}\frac{\delta}{\delta K_{\mu\nu}}
-\bar{c}_\mu\frac{\delta}{\delta \bar{c}_\mu}
-b_\mu\frac{\delta}{\delta b_\mu})\Gamma_{|s=1} \\
            =-c_3\kappa^{-2}[\int\sqrt{-g}R]^4_4\cdot\Gamma_{|s=1}
\end{multline}
Here the lhs has been chosen such that it is symmetric wrt ST (which can be checked most easily on ST for connected Green functions), whereas the rhs just follows from the action principle:
The BRST invariants $\int\sqrt{-g}(R^{\mu\nu}R_{\mu\nu}, R^2)$ are invariant under rigid Weyl, whereas the EH term is not.\\

When replacing the hard EH term by its soft version via appropriate Zimmermann identities (ZI) we obtain
\begin{multline}\label{rgWlII}
	W^{\rm W}_{\rm rig}\Gamma_{|s=1} =-2\sigma\lbrace
     \hat{\alpha} c_3\kappa^{-2}[\int\sqrt{-g}R]^3_3\cdot\Gamma_{|s=1}
	\\  +[\int\sqrt{-g}(u_1 R^{\mu\nu}R_{\mu\nu}+u_2 R^2)]^4_4
    +u_h\mathcal{N}_H+u_c\mathcal{N}_L\rbrace\cdot \Gamma_{|s=1}.
\end{multline}
Here $\hat{\alpha}=1+O(\hbar), u=O(\hbar)$, i.e.\ we have a soft term on the rhs which starts with the tree approximation, as desired.
Due to 
\begin{align}\label{psbCS}
	u_1\int\sqrt{-g}R^{\mu\nu}R_{\mu\nu}\cdot\Gamma=
	       u_1\frac{\partial}{\partial c_1}\Gamma\qquad
	u_2\int(\sqrt{-g}R^2)\cdot\Gamma=
               u_2\frac{\partial}{\partial c_2}\Gamma
\end{align}
the hard breaking terms going with $u_1$ and $u_2$ correspond directly to the 
conventional $\beta^{\rm CS}$-function terms of a CS equation, hence 
(\ref{rgWlII}) represents an adequate replacement of it:
\begin{align}\label{cs}
	\frac{-1}{2\sigma}W^{\rm W}_{\rm rig}\Gamma_{|s=1}
	-(u_1\frac{\partial}{\partial c_1} 
	+u_2\frac{\partial}{\partial c_2}
	+u_h\mathcal{N}_H+u_c\mathcal{N}_L)\cdot \Gamma_{|s=1}
= \hat{\alpha} c_3\kappa^{-2}[\int\sqrt{-g}R]^3_3\cdot\Gamma_{|s=1} \,.
\end{align}
From \cite{pottel2021perturbative} one should recall that the derivatives $\partial_{c_1}, \partial_{c_2}$ and the soft insertion on the rhs are symmetric wrt ST.\\

One can gain some more insight into the present case by going over to connected Green's functions via Legendre transformation and transition to general Green's functions 
\begin{multline}\label{rgWZ}
\frac{-1}{2\sigma}W^{\rm W}_{\rm rig}(\underbar{J})Z(\underbar{J})_{|s=1}
+(u_1[\int\sqrt{-g}R^{\mu\nu}]^4_4 
+u_2[\int\sqrt{-g}R^2]^4_4 
	-u_h\mathcal{N}_H-u_c\mathcal{N}_L)\cdot Z_{|s=1} \\
= \hat{\alpha} c_3\kappa^{-2}[\int\sqrt{-g}R]^3_3\cdot Z_{|s=1} \,,
\end{multline}
the subsequent projection (s.b.(\ref{prjfrl})), where $\mathcal{N}$'s vanish under the application of 
\begin{align}\label{eq:SMatrixSigma}	
	:\Sigma: \equiv \exp{\left\{\int dxdy\, \Phi_{\rm in}(x)K(x-y)z^{-1}
	\frac{\delta}{\delta\underbar{J}}\right\}}  
\end{align}	
which projects to the $S$-matrix,
whereas all other insertions have become operators
\be	([Q^{\rm W}_{\rm rig},S^{\rm Op}])
	-(u_1[\int\sqrt{-g}R^{\mu\nu}R_{\mu\nu}]^4_4)^{\rm Op} 
	-(u_2[\int\sqrt{-g}R^2]^4_4)^{\rm Op} 
	= \hat{\alpha} c_3\kappa^{-2}([\int\sqrt{-g}R]^3_3)^{\rm Op} \,.
\ee
$Q^{\rm W}_{\rm rig}$ denotes the charge operator of the rigid Weyl transformation.
The result can be interpreted as refering eventually to the physical
quartet states (again, for details s.\ Subsection \ref{sec:projectionEH}). \\ 
Hence the rigid Weyl tranformations are not a symmetry of the $S$-matrix: as
expected they are violated by the soft contribution originating from EH (rhs), 
and further by hard anomalies going with the higher derivative terms.
Obviously one can interpret those as providing the operators
$\int \sqrt{-g}(R^{\mu\nu}, R^2)$ with anomalous Weyl weights given by the
coefficients $u$. In the deep Euclidian limit $t\rightarrow \infty$ the
soft rhs vanishes.
One is then tempted to compare with the RG equation, s.b. (\ref{spRG}). However
there is no obvious relation between the coefficients $u$ and $\beta$, respectively.
These two equations refer to different physical issues.

\subsection{Finiteness in Landau gauge: $\gamma^{\rm RG}_c=\gamma^{\rm RG}_h=
\beta_3=0$}
It is well-known \cite{Blasi:1990xz} that in ordinary pure Yang-Mills theory the anomalous dimension of the vector field vanishes as well as that of the Faddeev-Popov ghost $c$ when working in Landau gauge.
Here we shall show that the analogue is true in the present model for the fields $h^{\mu\nu}$ and $c^\mu$.
This will then imply that $\beta_3$ also vanishes.
Obviously a tremendous impact on the model at hand.\\
In order to see this one starts from the integrated antighost equation of motion
\begin{align}\label{aghst}
	\int\frac{\delta\Gamma}{\delta c^\mu}\equiv& 
\int(\frac{\delta\Gamma_{\phi\pi}}{\delta c^\mu}
	+\frac{\delta\Gamma_{\rm e.f.}}{\delta c^\mu})\\ =
&-\kappa\int(\frac{1}{2}D^{\mu\nu}_\rho
        (\partial_\mu\bar{c}_\nu+\partial_\nu\bar{c}_\mu)
+K_{\mu\nu} D^{\mu\nu}_\rho+L_\lambda\partial_\rho c^\lambda
                         +\partial_\lambda(L_\rho c^\lambda))\\ 
	=&-\kappa\int(\frac{1}{2}D^{\mu\nu}
        (\partial_\mu\bar{c}_\nu+\partial_\nu\bar{c}_\mu)
-K_{\mu\nu} \partial_\rho h^{\mu\nu}+L_\lambda\partial_\rho c^\lambda)
\end{align}
and combines it with the gauge condition to form
\begin{align}\label{fntnc}
\bar{\mathcal{G}}\Gamma\equiv \int(\frac{\delta\Gamma}{\delta c^\rho}
+\kappa\partial_\rho\bar{c}_\lambda\frac{\delta\Gamma}{\delta b_\lambda})
= \int\kappa(K_{\mu\nu}\partial_\rho h^{\mu\nu} 
                            -L_\lambda\partial_\rho c^\lambda).
\end{align}
Since this expression is linear in the quantized fields it can be naively extended to all orders.
It is important to note that this equation not just follows in the tree approximation, but can be proved to hold to all orders of perturbation theory, s.\
Appendix \ref{sec:app_anti_ghost}.\\
Potential counterterms, one could have been obliged to add, must be independent of $b_\mu$, could depend on $\bar{c}$ only via $H_{\mu\nu}$ and satisfy the ghost equation
\begin{align}\label{ghqt}
\kappa\frac{\delta\bar{\Gamma}}{\delta\bar{c}_\mu}
+\partial_\lambda\frac{\delta\bar{\Gamma}}{\delta K_{\mu\lambda}}=0.
\end{align}
The candidates for this we have discussed in \cite[eq.\ (258)]{pottel2021perturbative}.
\begin{align}\label{ssrtL} 
\Delta_L\cdot\Gamma=f_L(\alpha_0)\mathcal{N}_L\Gamma-\chi 
       f^\prime_L\int Lc=-\mathcal{B}_{\bar{\Gamma}}(f_L\int Lc)\qquad
\mathcal{N}_L\equiv \int c\frac{\delta}{\delta c}
	            -\int L\frac{\delta}{\delta L}
\end{align}
In Landau gauge $\alpha_0=0, \chi=0$, hence $f_L$ is a number.
To be satisfied is (\ref{fntnc}), but 
\begin{align}\label{chck1}
\Delta_L\cdot\Gamma=-\mathcal{B}_{\bar{\Gamma}}(\int Lc)=
  \kappa\int L_\rho c^\lambda\partial_\lambda c^\rho + L{\rm -independent},
 \end{align}
would contribute a term 
$\int L_\lambda\partial_\mu c^\lambda$ to the rhs of (\ref{fntnc}),
which would, however, change the coefficient of the term already present. This
is forbidden, hence this term is excluded as a counterterm anb the external
field $L$ is not renormalized: $\gamma_c=0$, \cite[eq.\ (266)]{pottel2021perturbative}. The next
candidate is $\Delta_H$.
\begin{align}\label{ssrtH} 
\Delta_H\cdot\Gamma=f_H\mathcal{N}_K\Gamma
 +\chi f^\prime_H(\alpha_0)(\int(Kh
      -\bar{c}\frac{\delta\Gamma_{\rm class}}{\delta b}) 
              +2\alpha_0\frac{\partial}{\partial\chi}\Gamma_{\rm class}
	 \qquad     \mathcal{N}_K\equiv N_h-N_K-N_b
\end{align}
Again, in Landau gauge $\alpha_0=0, \chi=0$, $f_H$ is a number.
\be\label{ssrT2}
\Delta_H = f_H\int(-\kappa K_{\mu\nu}D^{\mu\nu}_\rho c^\rho)
\ee
This would contribute a term
$\int \kappa K_{\mu\nu}\partial_\rho h^{\mu\nu}$ to the rhs of (\ref{fntnc}), again a term which is 
already present and
whose coefficient must not be changed. So, this counterterm, too, is forbidden.
The field $h^{\mu\nu}$ is not renormalized: $\gamma_h$ vanishes \cite[eqs.\ (298), (300), (302)]{pottel2021perturbative}.
From $\gamma_c=\gamma_h=0$ follows $\beta_3=0$, \cite[eq.\ (303)]{pottel2021perturbative}.
This is the result advertised in the heading of the present section.\\

Several remarks are in order.
One can rather easily check in one-loop order that the diagrams contributing 
to $\Gamma_{Kc}$, resp.\ $\Gamma_{Lcc}$ are finite.
That they, in fact, vanish is not so obvious, however follows from the
preceding arguments.\\
The vanishing of $\beta_3$ is at variance  with results of the literature
of which we quote only \cite{AvBa}, \cite{NaAn},\cite{Ni}.
Our result is based on using the field $b_\mu=s\bar{c}_\mu$
for gauge fixing. This renders $s$ (and subsequently 
$\mathcal{B}_{\bar{\Gamma}})$ 
nilpotent and permits to construct Kugo-Ojima \cite{KuOj} quartets which 
define the
physical Hilbert space (s.\ Section \ref{sec:projectionEH}) in a clear cut fashion. 
The separation $\Gamma=\bar{\Gamma}+\Gamma_{\rm gf}$ is precisely in line with this and just
disentangles gauge fixing from contributions which are independent of it.
A concrete example is provided by eq.\ (\ref{kpp}) in the next section.
That the differential operator $\kappa\partial_\kappa$ has this $s$-invariant
extension originates from fixing the $h^{\mu\nu}$-field amplitude via
the external field $K_{\mu\nu}$, which itself is related to the separation
of $\Gamma$.
If one does not employ the field $b_\mu$ and these normalization conditions
this separation is virtually 
impossible. It is extremely difficult to disantangle the renormalization
of coupling $c_3$ from renormalization of the field amplitude for $h^{\mu\nu}$.
Historically there was a heated discussion on these points, just because 
also the proof of unitarity depends on it. Only with this technical tool 
provided by the
$b$ field one could master that problem and construct the physical
state space via Kugo-Ojima quartets. Becchi-Rouet-Stora did not use it in their first papers, 
but 
eventually adopted it. One may compare the original paper \cite{becchi1976renormalization}
with the Les Houches lectures of Becchi \cite{Becchi:1985bd}, where as concrete 
example the
same Yang-Mills theory with complete breakdown of the internal symmetry has
been discussed.\\
In the present context the additional problem is to separate the 
renormalization of $c_3$ and $\kappa$. Seduced by use of an invariant
renormalization scheme this has not been considered in the past. Our
scheme is not invariant. Hence one is forced to study in detail the origin
and effect of all possible counterterms, not just of invariant ones. This 
is admittedly cumbersome, but it is the best possible control of all 
possible relations in the model. Furthermore, in the literature one mostly
expanded also in $\kappa$ which however changes $\beta$ functions in a
drastic manner. We return to this point in the next section.

\subsection{The RG equation} \label{sec:rgequation}
We first state the relevant symmetric differential equations.
The symmetric $\kappa$ equation reads
\begin{align}\label{kpp}
		(-\kappa\partial_\kappa-2c_3\partial_{c_3}
	      -(N_b-2\alpha_0\partial_{\alpha_0})+N_K+N_L)\Gamma_{|s=1}=0,
\end{align}
hence this is a naive equation: no quantum corrections are associated with the 
differential operators. In particular this means that the combination
$c_3\kappa^{-2}$ does not give rise to a RG-running quantity, just because
there is no $\beta$-function like contribution in the equation.

The RG-equation is given by
\begin{align}\label{rg}
    (\mu\partial_\mu
     +\beta^{\rm RG}_3c_3\frac{\partial}{\partial {c_3}}
     +\beta^{\rm RG}_2c_2\frac{\partial}{\partial {c_2}}
     +\beta^{\rm RG}_1c_1\frac{\partial}{\partial {c_1}}
    -\gamma^{\rm RG}_h\mathcal{N}_H+\gamma^{\rm RG}_c\mathcal{N}_L)
                        \Gamma_{|s=1}=0 \, .
\end{align}
Their consistency results into
\begin{eqnarray}\label{cstkrg}
(\kappa\partial_\kappa+2c_3\partial_{c_3})\beta^{\rm RG}_i&=&0\quad i=1,2,3\\
(\kappa\partial_\kappa+2c_3\partial_{c_3})\gamma^{\rm RG}_i&=&0\quad i=c,h \, .
\end{eqnarray}
For the $h$-two-point function this reduces to
\begin{align}\label{2ptkpprg}
	(\kappa\partial_\kappa+2c_3\partial_{c_3})\Gamma_{{hh}_{|s=1}} & = 0\\
	\mu\partial_\mu\Gamma_{{hh}_{|s=1}} & =
     (-\beta^{\rm RG}_3c_3\frac{\partial}{\partial c_3}
     -\beta^{\rm RG}_2c_2\frac{\partial}{\partial c_2}
     -\beta^{\rm RG}_1c_1\frac{\partial}{\partial c_1}
	+2\gamma^{\rm RG}_h)\Gamma_{{hh}_{|s=1}} \, .
\end{align}
Taking into account that $\beta^{\rm RG}_3=\gamma^{\rm RG}_h=\gamma^{\rm RG}_c=0$ we end up with
\begin{align}\label{spRG} 
(\mu\partial_\mu+\beta^{\rm RG}_1 c_1\frac{\partial}{\partial c_1}
   +\beta_2 c_2\frac{\partial}{\partial c_2})\Gamma_{{hh}_{|s=1}} =0 \, .
\end{align}
We read off two obvious facts: due to their non-trivial $\beta$-functions the couplings $c_1,c_2$ can be extended to and used as running couplings in the solutions, whereas $c_3$ and $\kappa$ will not run.
Their values in the tree approximations do not have to be corrected in higher 
orders.\\
A side remark: As noted after $(\ref{rgWZ})$, this equation is close in form 
to the rigid Weyl 
identity, but describes a different physical situation.

This equation holds in particular also for their spin components $\gamma^{(r)}_{\rm KL}$.
Since the Landau gauge, $\alpha_0=0$, has been shown to be stable, the $(r=2,0; {\rm K=L=T})$ components do not become gauge parameter dependent.
Hence they depend on $p^2,\kappa$, and $\mu$.
The subtraction scheme guarantees a factor $p^2-m^2\equiv p^2-M^2(s-1)^2=p^2$ at $s=1$.
Hence (at $s=1$) $\gamma^{(r)}_{\rm TT}= \mathcal{O}(p^2)$. 
For the subsequent integration of the RG-equation it is convenient to separate the tree contributions from those of higher orders.
We therefore define
\begin{align}\label{nghhd}
	{\zeta}^{(r)}_{\rm TT}
    =\zeta^{(r)}_{\rm TT}
         (\frac{-p^2}{\mu^2},\frac{c_3\kappa^{-2}}{p^2},c_1,c_2) \, ,
\end{align}
which have vanishing naive $p$-dimension such that
\begin{align}\label{hhtgmm} 
\gamma^{(2)}_{{\rm TT}{|s=1}}=p^2c_3\kappa^{-2}
                  +p^2p^2 \zeta^{(2)}_{\rm TT}
\qquad \gamma^{(0)}_{{\rm TT}{|s=1}}=-2p^2c_3\kappa^{-2}
		  +p^2p^2 \zeta^{(0)}_{\rm TT}
\end{align}
Here the minus sign in the first variable of \eqref{nghhd} has been introduced, because the normalization conditions have been postulated at $p^2=-\mu^2$, a spacelike momentum; the $c_3$-dependence is in accordance with (\ref{2ptkpprg}).
We thus have to solve the equations ($r=1,2)$
\begin{align}\label{kpprggtrsv}
(\mu\partial_\mu
     +\beta^{\rm RG}_2c_2\frac{\partial}{\partial c_2}
     +\beta^{\rm RG}_1c_1\frac{\partial}{\partial c_1})
	\zeta^{(r)}_{\rm TT}
	=0,
\end{align}
if we wish to describe the implications of the RG equation for the respective vertex functions.
The resulting linear partial differential equations are homogeneous and can be solved via characteristics.
We define the variables
\begin{align}
t= \ln(-\frac{p^2}{\mu^2}) \qquad u=\frac{c_3\kappa^{-2}}{p^2} 
\quad {\rm i.e.} \quad ue^t=-\frac{c_3\kappa^{-2}}{\mu^2}
\end{align}
and the RG-equations become
\begin{align}
(\frac{\partial}{\partial t}
-\frac{1}{2}\beta_1(ue^t,g_1,g_2)g_1\frac{\partial}{\partial g_1}
-\frac{1}{2}\beta_2(ue^t,g_1,g_2)g_1\frac{\partial}{\partial g_1})
\zeta^{(r)}_{\rm TT}(e^t,e^tu,g_1,g_2)=0. 
\end{align}
Here the running couplings $g_i \gets c_i, i=1,2$ have to solve
\begin{align}\label{rcpls}
\frac{dg_i}{dt}=-\frac{g_i}{2}\beta^{\rm RG}_i\quad i=1,2\quad ({\textrm{no sum})}
\end{align}
(Prefactor and sign originate from using $\mu\partial_\mu$ instead of $\mu^2\partial_{\mu^2}$ and the explicit sign in the definition of $t$.)
Their solutions are given by
\begin{align}
g_i(t)=g_i(0)e^{-\frac{1}{2}\int^t_0 d\tau
		  \beta_i(ue^\tau,g_1(\tau),g_2(\tau))},
\end{align}
whereas
\begin{align}\label{tdpndu}
\frac{du}{dt}=0 \Rightarrow u(t,u_0,g_i(0))=u_0.
\end{align}
The separation of tree and higher order contributions, $(\ref{hhtgmm})$, in the original $\gamma^{(r)}_{\rm TT}$ determines the starting points of $\zeta$:
\begin{eqnarray}
\zeta^{(2)}_{\rm TT}(t,u_0,g_1,g_2)|_{t=0}&=&-g_1(0)
	     =-c_1\\
\zeta^{(0)}_{\rm TT}(t,u_0,g_1,g_2)|_{t=0}&=&3g_2(0)+g_1(0)
            =3c_2+c_1
\end{eqnarray}
A further restriction originates from the scheme, which fixes $\partial_{p^2}\gamma^{(2)}_{\rm TT}$ at $p^2=0,s=1$ and the fact that we did not admit counterterms which would change this.
Hence the dependence of $\zeta^{(r)}$ from $u=-\frac{c_3\kappa^{-2}}{p^2}$ is restricted to the value $u_0=-\frac{c_3\kappa^{-2}}{\mu^2}$.
Actually, this is just the content of (\ref{tdpndu}).
It fits to the fact, that $c_3\kappa^{-2}$ does not run. 
At this stage it is interesting to collect the results.
\begin{align}
	\gamma^{(2)}_{\rm TT}=&p^2p^2\zeta^{(2)}_{\rm TT}=
p^2c_3\kappa^{-2}+p^2p^2\zeta^{(2)}_{\rm TT}(e^t,u_0,g_1(t),g_2(t))\\
	\gamma^{(0)}_{\rm TT}=&p^2p^2\zeta^{(0)}_{\rm TT}=
-2p^2c_3\kappa^{-2}+p^2p^2\zeta^{(0)}_{\rm TT}(e^t,u_0,g_1(t),g_2(t))
\end{align}
The interpretation is as follows: the $c_3\kappa^{-2}$ terms are tree values which are not corrected by higher orders.
The $\zeta$ terms provide for the value $t=0$ the tree approximation contributions going with $c_1,c_2$. For $t> 0$ they comprise all higher order corrections expressed in terms of the running couplings.
One should note that these results hold at $s=1$, the physical value.
We underline, by repeating: the separation in tree, resp.\ higher order contributions in $\gamma^{(r)}_{\rm TT}, r=0,2$, $(\ref{hhtgmm})$, together with the non-renormalization of $\kappa^{-2}$ and $c_3$ just means that only the higher derivatives are responsible for the running of couplings, i.e.\ of $c_1$ and $c_2$.
In this respect the EH part is only a kind of spectator.

The common understanding of running couplings and their use in phenomenology
(QCD, electroweak standard model) is that inserting
them in place of a tree coupling at a given order in perturbation theory
``improves'' the results of that order, i.e.\ in some qualitative sense 
extends those to all orders. For the model under consideration,
in the literature mostly an expansion in terms of $\kappa^2$ has been performed.
However, we do not follow this path because the renormalization of the electroweak standard model teaches us an important lesson.
If one wants to ensure that there are poles for physical particles one has to use on-shell normalization conditions, which then requires the couplings to be realized as mass ratios.
This in turn implies that even their $\beta^{\rm CS}$-functions can no longer be
expanded as power series in the couplings, but only in the number of loops 
\cite{kraus1999callan}. (CS-$\beta$ functions are, as a rule, simpler than those
of the RG equation.)
The reason for this is the same as here: they are complicated polylogarithmic functions of the couplings.
Hence an expansion in terms of $\kappa^2$, here, may very well be misleading.
E.g.\ the pure fact that after using such an expansion the $\beta$-functions
come out as rational functions in the couplings is suspicious. Relying on
this outcome and concluding from there on the asymptotic behaviour seems
to be courageous.\\

\subsection{No massive higher order zeros}\label{sec:abshighzero}

Most important for the subsequent analysis is the understanding of the zeros of $\gamma^{(r)}_{\rm TT}$.
The ones at $p^2=0$ are fixed, guaranteed by the scheme and RG invariant: 
they are physical.
But the second zeros can not be continued to higher orders as we shall show now.\\
We consider the case $r=2$ up to and including one-loop.
\be\label{sp2gmm}
(\gamma^{(2)}_{\rm TT})^{(\le 1)}_{|s=1}=p^2c_3\kappa^{-2}-c_1p^2p^2
-c^{(1)}_1p^2p^2+p^2p^2(\zeta^{(2)}_{\rm TT})_{|\rm nt}
\ee
Here $c^{(1)}_1$ is the coefficient of the one-loop counterterm
to the invariant $\int\sqrt{-g}R^{\mu\nu}R_{\mu\nu}$; the subscript ``nt'' means
``non-trivial''
i.e.\ these are the contributions of the non-trivial diagrams in one-loop
order (the counterterm is pointlike, hence a trivial diagram). The first
zero at $p^2=0$ is obvious. We claim that
\be\label{2zgmm}
0=c_3\kappa^{-2}-c_1p^2-c^{(1)}_1p^2
            +p^2(\zeta^{(2)}_{\rm TT})_{|\rm nt}
\ee
has no solution for $p^2=\frac{c_3\kappa^{-2}}{c_1}$ and the counterterm 
coefficient with its value as given by the normalization condition for $c_1$
\begin{align}\label{cttc}
c^{(1)}_1&=\frac{1}{2}\frac{\partial}{\partial p^2}\frac{\partial}{\partial p^2}
(\gamma^{(2)}_{\rm TT})^{(1)}_{|p^2=-\mu^2,s=1}\\
	&=(\zeta^{(2)}_{\rm TT})^{(1)}_{|p^2=-\mu^2,s=1}
	+\lbrack (2p^2\partial_{p^2}
	+\frac{1}{2}p^2p^2\partial_{p^2}\partial{_p^2})
	                  (\zeta^{(2)}_{\rm TT})^{(1)}
	\rbrack_{|p^2=-\mu^2,s=1}.
\end{align}
Hence (\ref{2zgmm}) boils down to 
\be\label{xct}
\frac{c_3\kappa^{-2}}{c_1}(-c^{(1)}_1
 +(\zeta^{(2)}_{\rm TT})_{|\rm nt})=0,
\ee
with the arguments of $\zeta$ being 
$(-\frac{p^2}{\mu^2},\frac{c_3\kappa^{-2}}{p^2},c_1,c_2) \rightarrow  
(-\frac{c_3\kappa^{-2}}{(c_1\mu^2)},c_1,c_1,c_2) $. 
More explicitly
\begin{multline}\label{ctdt}
-\left( (\zeta^{(2)}_{\rm TT})^{(1)}
	(1,-\frac{c_3\kappa^{-2}}{c_1\mu^2},c_1,c_2)
	+\lbrack (-2\mu^2\partial_{p^2}
	+\frac{1}{2}\mu^2\mu^2\partial_{p^2}\partial{_p^2})
	              (\zeta^{(2)}_{\rm TT})^{(1)}
	\rbrack_{|p^2=-\mu^2,s=1}\right)_{|\rm nt} \\
+\left(\zeta^{(2)}_{\rm TT}
	(-\frac{c_3\kappa^{-2}}{c_1\mu^2},c_1,c_1,c_2)\right)_{|\rm nt} = 0,
\end{multline}
all taken at $s=1$. (In this explicit form also the first bracket refers to
the non-trivial diagrams.) It is to be noted that for the $\zeta$ in the 
first line
the  $p^2$-argument is at an unphysical value, whereas for the $\zeta$ in
the second line it is at a physical point. Therefor
this equation can not be satisfied. Hence beginning with one loop the
respective propagator, $\langle hh \rangle^{(2)}_{\rm TT}$,  has no second pole. Obviously, this is also true for the case $r=0$.

In the subsequent sections we shall determine the fields and respective 
propagators which can be associated with the two zeros of the tree 
approximation: 
the field belonging to the first zero generates massless particles, the field 
associated with the second  does not generate particles.

\section{Part II: Theory formulated in terms of $\phi$ and $\Sigma$}

\subsection{Lagrange multiplier form of the bilinear action}
In this section we search for free fields which belong to the partial
fractions of the double poles in the propagators $\langle hh \rangle$ as given in \cite{pottel2021perturbative}.
This tree approximation result can not be extended to higher orders
as we have shown in the preceding subsection. This will lead us to an 
interpretation and conclusion on the fate of the different fields involved.

Starting point is the decomposition of the $h^{\mu\nu}$ propagators into partial fractions which have only simple poles, as presented in \cite[eqs. 
$C_1,C_2$]{pottel2021perturbative}.
We are looking for fields $\phi^{\mu\nu}$ and $\Sigma^{\mu\nu}$ whose bilinear terms in the action just yield these simple pole propagators: $\phi$ the massless, $\Sigma$ the massive ones.
With this aim in mind one decomposes the field-bilinear part of the classical invariants of EH + hds with the help of a Lagrange multiplier $Z^{\mu\nu}$ such that only second derivatives of $h^{\mu\nu}$ and $Z^{\mu\nu},\Sigma^{\mu\nu}$ respectively show up.
Since in \cite{stelle1978classical} in an analogous context this problem has been solved we can proceed the other way round: we start from
\begin{equation}\label{hdcpt}
h^{\mu\nu}=\phi^{\mu\nu}+\Sigma^{\mu\nu}\qquad
Z^{\mu\nu}=\phi^{\mu\nu}-\Sigma^{\mu\nu}
\end{equation}
as desired field decomposition and from
\begin{align}\label{dcpta}
	\Gamma(\phi)&=\Gamma_{\rm EH}(\phi)\\
	&=\frac{\tilde{c}_3\kappa^{-2}}{4}\int(
-\phi^{\mu\nu}\Box\phi_{\mu\nu}+\phi^\rho_{\phantom{\rho}\rho}\Box\phi^\sigma_{\phantom{\sigma}\sigma}
-2\phi^{\mu\nu}\partial_\mu\partial_\nu\phi^\lambda_{\phantom{\lambda}\lambda}
	+2\phi^{\mu\nu}\partial_\rho\partial_\nu\phi^\rho_{\phantom{\mu}\mu}) \nonumber
\end{align}
\begin{align}\label{dcpta2}
	\Gamma(\Sigma)&=-\Gamma_{\rm EH}(\Sigma)+\Gamma_{\rm mass}(\Sigma)\\
	&=\frac{\tilde{c}_3\kappa^{-2}}{4}\int(
\Sigma^{\mu\nu}\Box\Sigma_{\mu\nu}-\Sigma^\rho_{\phantom{\rho}\rho}\Box\Sigma^\sigma_{\phantom{\sigma}\sigma}
+2\Sigma^{\mu\nu}\partial_\mu\partial_\nu\Sigma^\lambda_{\phantom{\lambda}\lambda}
	-2\Sigma^{\mu\nu}\partial_\rho\partial_\nu\Sigma^\rho_{\phantom{\mu}\mu} \nonumber    \\
	&\phantom{\frac{c_3\kappa^{-2}}{4}\int)}
        +a_2\Sigma^{\mu\nu}\Sigma_{\mu\nu}
	+a_0(\Sigma^\lambda_{\phantom{\lambda}\lambda})^2)   \nonumber  
\end{align}
as desired bilinear action in order to identify at a convenient stage in our conventions the mass $a_2,a_0$ and coupling $\tilde{c}_3\kappa^{-2}$ parameters. 
The relative minus sign of the two actions just represents the negative residue sign of massive propagators in \cite[eqs. $C_1,C_2$]{pottel2021perturbative}.
As an aside we note that the mass term is not of Fierz-Pauli type, since it will turn out that $a_2+a_0\not=0$.
Hence it contains some spin $0$ component.
(The Fierz-Pauli condition $a_2+a_0=0$ would remove within $\gamma^{(0)}_{\rm TT}$ the second zero, hence ruin UV convergence.)\\
For the subsequent treatment we give here the actions in projector form. (The
projectors can be found in \cite[App.\ A]{pottel2021perturbative})
\begin{align}\label{prjctfrm}
\begin{split}
	\Gamma_{\rm EH}(\phi)=\frac{\tilde{c}_3\kappa^{-2}}{4}\int(
&-\phi^{\mu\nu}\Box(P^{(2)}_{\rm TT}+P^{(1)}_{\rm SS}
                  +P^{(0)}_{\rm TT}+P^{(0)}_{\rm WW})\phi^{\rho\sigma}\\
&+\phi^{\mu\nu}\Box(3P^{(0)}_{\rm TT}+\sqrt{3}(P^{(0)}_{\rm TW}
		  +P^{(0)}_{\rm WT})+P^{(0)}_{\rm WW})\phi^{\rho\sigma}\\
&-\phi^{\mu\nu}\Box(\sqrt{3}(P^{(0)}_{\rm TW}+P^{(0)}_{\rm WT})
	            +2P^{(0)}_{\rm WW})\phi^{\rho\sigma}\\
	&+\phi^{\mu\nu}\Box(P^{(1)}_{\rm SS}
		    +2P^{(0)}_{\rm WW})\phi^{\rho\sigma} 
\end{split}
\end{align}
\begin{align}\label{prjct2}
\begin{split}
	\Gamma(\Sigma)
	=\frac{\tilde{c}_3\kappa^{-2}}{4}\int(
	&\Sigma^{\mu\nu}(\Box+a_2)(P^{(2)}_{\rm TT}+P^{(1)}_{\rm SS}
                  +P^{(0)}_{\rm TT}+P^{(0)}_{\rm WW})\Sigma^{\rho\sigma}\\
 &-\Sigma^{\mu\nu}(-\Box+a_0)(3P^{(0)}_{\rm TT}+\sqrt{3}(P^{(0)}_{\rm TW}
		  +P^{(0)}_{\rm WT})+P^{(0)}_{\rm WW})\Sigma^{\rho\sigma}\\
&+\Sigma^{\mu\nu}\Box(\sqrt{3}(P^{(0)}_{\rm TW}+P^{(0)}_{\rm WT})
	            +2P^{(0)}_{\rm WW})\Sigma^{\rho\sigma}\\
&-\Sigma^{\mu\nu}\Box(P^{(1)}_{\rm SS}
		    +2P^{(0)}_{\rm WW})\Sigma^{\rho\sigma})
\end{split}
\end{align}
In the next step we replace the fields: $\phi=h+Z, \Sigma=h-Z$ and go over to a total action
\begin{align}\label{tctn}
\Gamma(\phi)+\Gamma(\Sigma)=\Gamma_{\rm total}\rightarrow \Gamma(h,Z)
\end{align}
We find 
\begin{align}\label{hZfctn}
\begin{split}
	\Gamma_{\rm total}(h,Z)=\frac{\tilde{c}_3\kappa^{-2}}{4}\int(
	&-2h(P^{(2)}_{\rm TT}(2\Box+a_2)+P^{(1)}_{\rm SS}a_2
	 +P^{(0)}_{\rm TT}(-4\Box+a_2+3a_0)\\
	& +\sqrt{3}(P^{(0)}_{\rm TW}
	 +P^{(0)}_{\rm WT})a_0+P^{(0)}_{\rm WW}(a_2+a_0))Z\\
        &+h(P^{(2)}_{\rm TT}a_2+P^{(1)}_{\rm SS}a_2+P^{(0)}_{\rm TT}(a_2+3a_0)\\
	&+\sqrt{3}P^{(0)}_{\rm TW}+P^{(0)}_{rm WT}a_0+P^{(0)}_{\rm WW}(a_2+a_0)
	 )h\\
        &+Z(P^{(2)}_{\rm TT}a_2+P^{(1)}_{\rm SS}a_2+P^{(0)}_{\rm TT}(a_2+3a_0)\\
	&+\sqrt{3}P^{(0)}_{\rm TW}+P^{(0)}_{rm WT}a_0+P^{(0)}_{\rm WW}(a_2+a_0)
	 )Z \,.
\end{split}
\end{align}
This action has the desired structure $\int(h\mathcal{D}_{hZ}Z+hMh+Z\mathcal{D}_{ZZ}Z)$ with $Z$ representing the Lagrange multiplier field.
This explicit form has not been presented in \cite{stelle1978classical}.\\
The final form $\Gamma(h)$, which can be compared with EH+hds, is now obtained by eliminating $Z$ via its equation of motion
\begin{align}\label{qtnmZ}
\frac{\delta \Gamma}{\delta Z}(h,Z)=0.
\end{align}
One obtains
\begin{align}\label{rstZH}
\begin{split}
	(P^{(2)}_{\rm TT}a_2	
	+P^{(1)}_{\rm SS}a_2	
	+P^{(0)}_{\rm TT}(a_2+3a_0)	
	+\sqrt{3}(P^{(0)}_{\rm TW}+P^{(0)}_{\rm WT})a_0	
	+P^{(0)}_{\rm WW}(a_2+a_0))Z\\	
	=(P^{(2)}_{\rm TT}(2\Box +a_2) 	 
	+P^{(1)}_{\rm SS}a_2	
	+P^{(0)}_{\rm TT}(-4\Box+a_2+3a_0)\\	
	+\sqrt{3}(P^{(0)}_{\rm TW}+P^{(0)}_{\rm WT}a_0	
	+P^{(0)}_{\rm WW}(a_2+a_0))h	 \,.
\end{split}
\end{align}
Suitably projecting and  equating coefficients one can solve for $Z$ in terms of $h$.
Inserting into (\ref{hZfctn}) one arrives finally at
\begin{align}\label{frth}
	\Gamma_{\rm total}(h)=\frac{\tilde{c_3}\kappa^{-2}}{4}\int h
   (P^{(2)}_{\rm TT}(-\frac{4}{a_2}\Box(\Box+a_2))
	+P^{(0)}_{\rm TT}(-\frac{8}{a_2+3a_0}
	                \Box(\Box+(a_2+3a_0))))h \,.
\end{align}
This result permits identification of the parameters:
\begin{align}\label{prntfn}
	\frac{1}{4}\tilde{c}_3\kappa^{-2}=c_3\kappa^{-2} 
 \qquad             a_2=\frac{4c_3\kappa^{-2}}{c_1}
\qquad a_0=-\frac{3c_2+2c_1}{3c_1(3c_2+c_1)}c_3\kappa^{-2} \,.
\end{align}

\subsection{Propagators}
This section is not mandatory for the understanding and use of the Lagrange multiplier trick within EH + hds, but still instructive.
The fact that one can invert (to the propagators) gives the massive submodel some ``stand-alone'' properties which are important when considering cohomology (in the classical approximation), s.\ Sect.\ \ref{sec:cohomology}.\\

Eventually the field $\phi$ will be interpreted as a standard gravitational field for a standard EH, whereas $\Sigma$ will be considered as a kind of auxiliary field whose only purpose is to guarantee power counting renormalizability, a ``shadow'' field. 
Hence $\phi$ has the standard transformation properties under BRST
\begin{align}\label{trsfphi}
s\phi^{\mu\nu}=-\kappa(\partial^\mu c^\nu+\partial^\nu c^\mu +
\partial_\lambda c^\mu \phi^{\lambda\nu}+
\partial_\lambda c^\nu \phi^{\mu\lambda}
	-c^\lambda\partial_\lambda \phi^{\mu\nu}) 
\end{align}
and requires gauge fixing etc as usual (s. \cite{pottel2021perturbative}).
For the field $\Sigma$ we could  assign 
\begin{align}\label{trsfS}
s\Sigma =s_2\Sigma=-\kappa(\partial_\lambda c^\mu \Sigma^{\lambda\nu}
			  +\partial_\lambda c^\nu \Sigma^{\mu\lambda}
			  -c^\lambda\partial_\lambda \Sigma^{\mu\nu})
\end{align}
Since, however such a non-linear transformation would not survive when going on-shell we are allowed to assume $s\Sigma=0$ -- which we will choose.\\

$\phi$ has the standard propagators,
\begin{align}\label{prpsfi}
	\langle\phi\phi\rangle^{(2)}_{\rm TT}=&\frac{i}{\tilde{c}\kappa^{-2}p^2}\quad
	\langle\phi\phi\rangle^{(0)}_{\rm TT}=\frac{-i}{2\tilde{c}\kappa^{-2}p^2}\\
	\langle\phi\phi\rangle^{(1)}_{\rm SS}=&\frac{4i\alpha_0\kappa^2}{p^2}\quad
	\langle\phi\phi\rangle^{(0)}_{\rm WW}=\frac{4i\alpha_0\kappa^2}{p^2}\quad
	\langle\phi\phi\rangle^{(0)}_{\rm TW}=\langle\phi\phi\rangle^{(0)}_{\rm WT}=0\\
\langle b_\rho\phi_{\mu\nu}\rangle=&\frac{\kappa}{p^2}(p_\mu\theta_{\nu\rho}
	                                +p_\nu\theta_{\mu\rho} 
					+p_\rho\omega_{\mu\nu})\qquad\quad
                \langle b_\rho b_\sigma \rangle=0	                     
\end{align}
and the standard Fock space of a massless spin 2 field.
Like in the case of an abelian vector field with ad hoc added mass term 
\cite{Bogolyubov:1980nc}, here too, $\Sigma$ has a propagator which can be obtained without further ado.
It is determined by the postulate
\begin{align}\label{nvrsS}
\Gamma_{\Sigma^{\mu\nu}\Sigma^{\kappa\lambda}}
G_{\Sigma^{\lambda\kappa}\Sigma^{\rho\sigma}}=
\frac{i}{2}(\delta_\mu^\rho\delta_\nu^\sigma+\delta_\mu^\sigma\delta_\nu^\rho)
=i(P^{(2)}_{\rm TT}+P^{(1)}_{\rm SS}+P^{(0)}_{\rm TT} 
                  +P^{(0)}_{\rm WW})^{\rho\sigma}_{\mu\nu}
\end{align}
Like the actions we also expand the propagators in projected form
\begin{align}\label{vrlldfpr}
G(\Sigma)^{\rho\sigma}_{\mu\nu} =\frac{\tilde{c}_3\kappa^{-2}}{4}
\sum_{r,K,L}
\langle\Sigma\Sigma\rangle^{(r)}_{\rm K,L}P^{{(r)}{\rho\sigma}}_{{\rm KL}{\mu\nu}}
\end{align}
and insert into (\ref{vrlldfpr}).
One easily finds
\begin{align}\label{2TT}
	\langle\Sigma\Sigma\rangle^{(2)}_{\rm TT}=\frac{i}{\Box+a_2}\delta(x-y)
\qquad  \langle\Sigma\Sigma\rangle^{(1)}_{\rm SS}=\frac{i}{a_2}\delta(x-y)
\end{align}
We obtain the other propagators in two steps, since the projectors with $r=0$ are not all orthogonal to each other.
We first group the terms as follows
\begin{align}\label{rgrtrs}
\begin{split}
 P_{\rm TT}(-i+\gamma_{\rm TT}\langle\Sigma\Sigma\rangle_{\rm TT}
	   &+\gamma_{\rm tf}\langle\Sigma\Sigma\rangle_{\rm tf})
+P_{\rm WW}(-i+\gamma_{\rm WW}\langle\Sigma\Sigma\rangle_{\rm WW}
	+\gamma_{\rm tf}\langle\Sigma\Sigma\rangle_{\rm tf})\\
&=P_{\rm TW}(-\gamma_{\rm TT}\langle\Sigma\Sigma\rangle_{\rm tf}
            -\gamma_{\rm TT}\langle\Sigma\Sigma\rangle_{\rm WW}
	    -\gamma_{\rm tf}\langle\Sigma\Sigma\rangle_{\rm WW})\\
& \quad +P_{\rm WT}(-\gamma_{\rm WW}\langle\Sigma\Sigma\rangle_{\rm tf}
            -\gamma_{\rm WW}\langle\Sigma\Sigma\rangle_{\rm TT}
	    -\gamma_{\rm tf}\langle\Sigma\Sigma\rangle_{\rm TT}).
\end{split}
\end{align}	    
(In order to simplify the writing we have omitted the superskripts $(0)$.
``tf'' means ``transfer'', $\gamma_{tf}=\sqrt{3}a_0$.)
Then we multiply the lhs with the lhs and the rhs with the rhs.
The resulting equation constitutes a kind of consistency condition which is automatically satisfied once we impose
\begin{eqnarray}\label{prpsTTWW}
	\langle\Sigma\Sigma\rangle^{(0)}_{TT}&=&\frac{i}{\gamma_{\rm TT}}
	           =\frac{i}{-2\Box+a_2+3a_0}\delta(x-y)\\
	\langle\Sigma\Sigma\rangle^{(0)}_{WW}&=&\frac{i}{\gamma_{\rm WW}}
	           =\frac{i}{a_2+a_0}\delta(x-y)
\end{eqnarray}
The equation (\ref{rgrtrs}) enforces finally
\begin{align}\label{prptf}
\langle\Sigma\Sigma\rangle^{(0)}_{\rm tf}=0.
\end{align}
The propagators $r=2,0; K=L=T$ are truely dynamic, the others are not.

\subsection{$s$-cohomology for $\phi$ and $\Sigma$ sectors}\label{sec:cohomology}

Looking at \eqref{dcpta}, \eqref{dcpta2}, (\ref{prjctfrm}) it is clear that one can endow the field $\phi$ with the standard transformation $s=s_0$.
The respective action $\Gamma(\phi)_2$ is invariant under it and can by cohomology be extended to a complete (in the expansion in the number of fields) $\Gamma(\phi)= \Gamma_{\rm EH}(\phi)$ invariant under the complete $s=s_0+s_1$, i.e.
\begin{align}\label{stehact}
\Gamma_{\rm EH}=\frac{c_3\kappa^{-2}}{4}\int\sqrt{-g}\,R(\phi) \,.
\end{align}
In order to exhaust all possibilities of $s$-covariance we admit now $s\Sigma=s_0 \Sigma$.
Within $\Gamma_2(\Sigma)$, \eqref{dcpta}, \eqref{dcpta2}, (\ref{prjctfrm}) the mass term is not invariant under $s=s_0$.
Invariance is, however, achieved on the mass shell, i.e.\ using the equation of motion for $\Sigma$. 
It is also obvious that the $\Sigma$ terms with (two) derivatives can be continued to all orders in the number of fields, hence accordingly symmetric under the complete $s=s_0+s_1$.
Together with the mass term, this constitutes (modulo overall sign) massive spin two \& some spin 0.
We recall that it is not of Fierz-Pauli type. \\
How does the hds version arise via cohomology?
Obviously the $\Sigma$-terms without derivatives (the mass terms) can be, if at all, only removed by going on-shell.
Hence one should not consider them any more when searching for hds.
Instead one observes that the general solution of
\begin{align}\label{xtehSgm}
s((-\Gamma_{\rm EH}(\Sigma))_2)=0
\end{align}
is given by 
\begin{align}\label{ehhds}
\Gamma =-\Gamma_{\rm EH}(\Sigma)+\int\sqrt{-g}(c_1R^{\mu\nu}R_{\mu\nu}+c_2R^2) \,.
\end{align}
The pure hds terms arise as a kind of  ``integration constants'' when 
performing cohomology with respect to $s$, just because they are allowed by 
power counting and are, of course, invariant.
The first term in (\ref{ehhds}) is the extension in the number of fields which we had obtained already.\\
The transition to our desired solution thus requires some additional argument apart from just cohomology.
We extend (\ref{stehact}) by demanding that $\phi$ gets replaced by
$\phi+\Sigma=h$.
And we extend hds of (\ref{ehhds}) by demanding the replacement
$\Sigma \rightarrow \Sigma+\phi$.
The reason for doing so is based on power counting and convergence: (\ref{stehact}) can not stand alone because it is not UV finite, neither is the massive model, and similarly the hds-terms of (\ref{ehhds}) alone are not IR convergent (all by power counting \cite{pottel2021perturbative}).
Only the combined terms, EH + hds, are convergent and only these combined arguments of $s$-cohomology and power counting lead to existing models.\\
The main reason for this round-about argument is just that it is non-trivial to realize the Lagrange multiplier construction in higher orders because of the non-linearity in terms of the fields involved.
In the tree approximation it would however go through and thus provide insight and justification for the discussion.

\subsection{Projection to Einstein-Hilbert} \label{sec:projectionEH}
We start from the model constructed to all orders in
\cite{pottel2021perturbative} in terms of the double pole field $h^{\mu\nu}$ 
and indicate now, how to identify the fields $\phi,\Sigma$ and their use 
within that given model.
The massive field $\Sigma$ will, beginning with one loop, no longer refer to 
the propagation of a stable particle: its propagator has at the very best a 
complex pole. (Due to the properties of the polylogs it could also have 
another singular character.)
Furthermore there are no parameters available which could in accord with the
$s$-symmetry protect the real part of the possible singularity from being 
shifted in higher orders and thus is not invariant under RG in accordance 
with $(\ref{2zgmm})$. A clear hint that it is unphysical, apart from its
negative norm properties in the tree approximation. We continue this discussion
after having described the projection procedure to the physical Hilbert space.

We identify the massless spin two field
$\phi^{\mu\nu}$ with the massless spin two graviton field, together with the 
fields $c$, $\bar{c}$, and $b$ as companions for building up the Kugo-Ojima \cite{KuOj} 
doublets. We can proceed for $\phi$ this way because it satisfies all 
requirements which one expects for such a field. Is has the correct 
covariance under $s$ and can in all respects be obtained from
\cite{pottel2021perturbative} by replacing $h^{\mu\nu}$ within 
$\Gamma_{\rm eff}=\Gamma^{\rm class}+\Gamma^{\rm counter}$
with $h^{\mu\nu}=\phi^{\mu\nu}+\Sigma^{\mu\nu}$ in the field expansion with the number of fields $n$ greater or equal to three.
(In particular, for the counterterms,too, $h=\phi+\Sigma\,$.)
In tree approximation bilinear terms and in gauge fixing, Faddeev-Popov, and 
external field terms, $h$ is simply replaced by $\phi$, whereas the field $\Sigma^{\mu\nu}$ comes along with the terms given in (\ref{dcpta2}) (upon replacing the mass parameters $a_2,a_0$ with their values given in (\ref{prntfn})).\\
The general Green's functions in terms of $h$ give rise to those of 
$\phi$ and $\Sigma$ for number of fields $n$ greater or equal to three by 
introducing respective sources $j_\phi$ and $j_\Sigma$, fitting to $h=\phi+\Sigma$.
We go over to the $S$-operator by applying a projector $:\!\Sigma\!: = :\!\exp{(Y)}\!:$, as
mentioned before in \eqref{eq:SMatrixSigma}. 
The free in-fields are related to their corresponding wave function operator
$K$ as follows:
\begin{align}\label{prjfrl}
\begin{split}
	\ln :\Sigma: = Y \qquad \quad \, &\\
	\equiv \int dxdy \Big( &
	\phi^{\mu\nu}(x)K^{\phi\phi}_{\mu\nu\rho\sigma}(x-y)z^{-1}
	\frac{\delta}{\delta j^{\phi}_{\rho\sigma}}(y)\\
	&+\phi^{\mu\nu}(x)K^{\phi b}_{\mu\nu\rho}(x-y)z^{-1}
	\frac{\delta}{\delta j^b_\rho}(y)\\
	&+b^\rho(x)K^{b\phi}_{\rho\alpha\beta}(x-y)z^{-1}	
	\frac{\delta}{\delta j^\phi_{\alpha\beta}}(y)\\
	&+b^\rho(x)K^{bb}_{\rho\sigma}(x-y)z^{-1}
	\frac{\delta}{\delta j^b_\sigma}(y)   \\
	&+\Sigma^{\mu\nu}(x)K^{\Sigma\Sigma}_{\mu\nu\rho\sigma}(x-y)z^{-1}
	\frac{\delta}{\delta j^{\Sigma}_{\rho\sigma}}(y)\\
	&+c^\rho(x)K^{c\bar{c}}_{\rho\sigma}(x-y)z^{-1}
	\frac{\delta}{\delta j^{\bar{c}}_\sigma}(y)\\
	&+\bar{c}^\rho(x)K^{\bar{c}c}_{\rho\sigma}(x-y)z^{-1}
	\frac{\delta}{\delta j^c_\sigma}(y) \Big)
\end{split}
\end{align}
The factors $z^{-1}$ stand for the inverse residue of the respective propagator.
(The reference to which one, we have suppressed for notational convenience.)
All fields are free in-fields and the projector $:\Sigma:$ projects down to their Fock spaces.

The Hilbert space for the $\phi$ quartets is defined in the following way
\cite{KuOj}.
Amongst the states $|\phi,c,\bar{c},b\rangle$, made up by the fields indicated, one 
selects those which are annihilated by the BRST-charge $Q$: $Q|\rm{phys}\rangle=0$.
Since $\Sigma$ is $s$-invariant their Fock space which contains negative 
norm states is still part of it. All of them build up the state 
$\mathcal{V}_{\rm phys}$.
Up to the $\Sigma$ subspace the norms of all other vectors is known
to be non-negative.\\
As far as the $\Sigma$ fields are concerned we now conclude that in higher
orders they are projected to zero. This is due to the fact 
that their original real poles in the tree 
approximation have been shifted on the real axis and into the complex $p^2$
plane -- a change which we could not prohibit via (symmetric) counterterms, 
because those are not available. They have been used for fixing the
symmetric invariants $\int \sqrt{-g}(c_1 R^{\mu\nu}R_{\mu\nu}+c_2 R^2)$.
In tree approximation there are, however, still nonvanishing contributions.
One might be tempted to put there ``by hand'' $c_1=c_2=0$,
with the argument that in tree approximation no higher derivatives are required. 
But this is in conflict with the solutions $g_i(t)$ of the RG-equation 
(\ref{rcpls}) which then vanish.\\
Hence one has to live with some loss of probability in tree approximation: 
All initial states made up
from $\phi$ which go into final states made up from $\Sigma$ prohibit
that positivity is realized. One can consult in this context the paper
\cite{InAb}, where (although with another aim in mind) explicitly
such processes have been studied and one can see that the higher derivatives
play already in tree approximation the important role of damping amplitudes.
Beginning with one-loop the $\Sigma$-states can no longer be excited as
outgoing states, hence there one has as final state the above described quartet 
states.\\

The states with vanishing norm form a linear subspace $\mathcal{V}_0$ of it 
and can thus be modded out.
The closure of this space forms the Hilbert space of the $\phi$-quartet 
subspace:
\begin{align}\label{Hspphi}
\mathcal{H}=\overline{\mathcal{V}_{\rm phys}/\mathcal{V}_0}
\end{align}
It implies $SS^\dagger=S^\dagger S=1$, because $\phi$ and its companions form
a quartet such that only their non-negative norm states survive in the 
$S$-matrix.
It is to be noted, however that still in internal lines of diagrams the field $\Sigma$ is present and plays its growth limiting role, since there $h=\phi+\Sigma$ and the propagators still have their UV-falloff with $(p^2)^{-2}$.\\
The internal lines consist of $\langle\phi\phi\rangle+\langle\Sigma\Sigma\rangle$ (+ other  members 
of the quartet). In the appendix we discuss how the optical theorem can be 
realized.

Hence the $\Sigma$ fields and their interactions are not irrelevant as far as physics is concerned.
The transition amplitudes of the $\phi$ fields amongst their quartet states will
in general depend on the couplings $c_1$ and $c_2$ and thereby exhibit the influence of
the ``shadow world'' spanned by $\Sigma$'s.

\section{Discussion and Conclusions} \label{sec:conclusions}
The present paper is based on the renormalization to all orders of the EH action + higher derivatives which we described in \cite{pottel2021perturbative}.
The upshot of that investigation was that using a spin two field $h^{\mu\nu}$ which has canonical dimension zero and a double pole propagator permits to derive (almost) all results one is interested in: the ST identity is established, which yields an invariant $S$-matrix, and parametric differential equations exist.
Redefinitions of the field $h$ as function of itself are also under control. \\
In the present paper we derived some interesting finiteness properties stemming from Landau gauge.
They simplify considerably the RG equation because they tell one that the EH coupling $c_3$ and the paramter $\kappa^{-2}$ do not run.
The Ward identity for rigid Weyl transformations which we also derived are the correct substitute for scaling, i.e.\ the CS equation, in a genuine flat space theory.\\

The only critical issue which had not been thoroughly discussed in \cite{pottel2021perturbative} was the unitarity of the $S$-matrix.
We proposed a projection down to a Hilbert space constructed out of Kugo-Ojima quartet states, but it was left open to which extent unitarity is indeed realized.\\
In order to clarify this issue we introduced here spin two single-pole fields $\phi^{\mu\nu}$ massless, and $\Sigma^{\mu\nu}$ massive, such that the original action in order two of $h$ is separated and diagonalized. 
They just realize separately the poles which $h$ posesses in the tree approximation.
Since the version in terms of $h$ had been completely discussed, one only had to find out which contributions must be attributed to which of the new fields.
In the interaction and counterterms this is just the sum $h=\phi+\Sigma\,$: hence all previously obtained results (perturbation theory, convergence and cohomology) can be carried over.
In this sum, in particular the UV and (off-shell) IR problems remain solved.
The bilinear terms are just what is needed to identify the state space: For $\phi,c,b,\bar{c}$ this is the Hilbert space as constructed out of Kugo-Ojima quartets.
For $\Sigma$ -- the massive field -- however, and this is the most important new
aspect of the present paper, it is a result of closer inspection, 
that its propagator has -- beginning with one-loop -- at most a (complex) 
pole with real part shifted away from the free theory value and not being
RG-invariant.
Hence the projection to the $S$-matrix yields zero beginning with one-loop. 
In the tree approximation however this will not happen, hence we find
a loss of probability caused by transitions of $\phi$'s to $\Sigma$'s.
A reparametrization by trading the two coupling parameters $c_1$ and $c_2$ against the
two mass paramters $a_2$ and $a_0$ would not be compatible with $s$-invariance and 
existence of an RG-equation and is thus forbidden.
This outcome then ensures unitarity of the $S$-matrix, just because the 
Hilbert space of the theory is spanned by the quartets of $\phi$ alone.

In general, Green's functions, $S$-matrix and operator matrix elements will 
depend on the couplings $c_1$ and $c_2$.
Hence the higher derivatives are a substantial element of the theory and 
in this sense there does not exist a quantized gravity model based on EH 
alone.

\begin{appendix}
\section{Appendix}
\subsection{Proof of antighost equation}\label{sec:app_anti_ghost}
We give a proof of the antighost equation to all orders of perturbation theory.
Following closely \cite{Blasi:1990xz} we insert in the rhs of \eqref{fntnc} a correction term $\hat{\Delta}_\rho$
\be
\bar{\mathcal{G}}_\rho\Gamma=\Delta_\rho^{\rm G}+\hat{\Delta}_\rho
\ee
Here $\Delta^{\rm G}_\rho=\int(\kappa K_{\mu\nu}\partial_\rho g^{\mu\nu}
                  -\kappa L_\lambda\partial_\rho c^\lambda)$
represents the tree contribution, whereas $\hat{\Delta}_\rho$ collects higher
order corrections which are in one-loop order just local terms.
Then we apply the constraints provided by the algebraic relations
\begin{align}
\lbrack \frac{\delta}{\delta b},\bar{\mathcal{G}}\rbrack=&0\\	
	\lbrace\bar{\mathcal{G}}_\rho,\bar{\mathcal{G}}_\sigma\rbrace=&0\\
	\lbrace\bar{\mathcal{G}_\rho},\mathcal{G}^\mu\rbrace=
	       &\kappa\partial_\rho\frac{\delta}{\delta b_\mu}\\
       \bar{\mathcal{G}}_\rho\mathcal{S}(F)
		+S_F(\bar{\mathcal{G}}_\rho F)=&0
\end{align}
for any functional $F$. One finds the following restrictions on 
$\hat{\Delta}_\rho$
\begin{align}
	\frac{\delta}{\delta b}\hat{\Delta}_\rho=&0 \label{A6}\\
	\mathcal{G}^\mu(x)\hat{\Delta}_\rho=&0 \label{A7}\\
	\bar{\mathcal{G}}_\rho \hat{\Delta}_\sigma
	+\bar{\mathcal{G}}_\sigma \hat{\Delta}_\rho=&0 \label{A8}\\
	\mathcal{S}_\Gamma \hat{\Delta}_\rho=&0\label{A9}
\end{align}
$\hat{\Delta}_\rho$ has ghost number +1 and dimension 4. Hence a basis
which satisfies the constraints $(\ref{A6},\ref{A7})$ is
provided by 
$\int\kappa L_\lambda\partial_\rho c^\lambda,\int H_{\mu\nu}h^{\mu\nu}$.
Constraint $(\ref{A8})$ holds then automatically. An explicit check
of constraint $(\ref{A9})$ shows that it can not be satisfied, hence
$\hat{\Delta}_\rho=0$ at first order in $\hbar$, by induction then to all
orders. So, the antighost equation is proven to all orders.\\

\subsection{Optical Theorem} \label{sec:optictheorem}
The derivation of the optical theorem  starts with the definition of
a transition operator $T$ and the unitarity relation for the $S$-operator.
\be\label{TU}
S=1+iT \qquad SS^\dagger=S^\dagger S=1 \qquad -i(T-T^\dagger)= TT^\dagger
\ee
Inserting the completeness relation $\Sigma_n |n\rangle \langle n|=I$
one arrives at
\be\label{pstmu}
-i(T-T^\dagger)=\Sigma_n T|n\rangle\langle n|T^\dagger.
\ee
Taking suitable matrix elements one obtains interesting bounds.

In the presence of states which have negative metric and are non-diagonal
the relations, (\ref{TU}), (\ref{pstmu}) are to be changed \cite[(3.2.32), (3.2.35)]{TKB}.
The complete $S$-matrix describes also the scattering of fields which 
lead to states with negative or vanishing norm. 
Unitarity becomes ``pseudo''-unitarity and the completeness relation
takes care of negative norm and non-diagonality by a metric 
$\eta_{\alpha\beta}$
\be\label{PTU}
\Sigma_{\gamma,\delta} S_{\beta\gamma}\eta^{-1}_{\gamma\delta}
               S^\dagger_{\delta\alpha}= \eta_{\beta\alpha}
\ee
We now consider elements of the physical Hilbert space (\ref{Hspphi}).
States containing $|\Sigma...>$ are still present as vectors in 
$\mathcal{V}_{\rm phys}$, because the 
$\Sigma$-field is invariant under $Q^{\rm BRST}$, hence also their Fock 
states are. However beginning with one-loop
they are projected to zero under $:\Sigma:$ as has been explained before.
In tree approximation they are, however still present.
Elements contained in $\mathcal{V}_0$ are orthogonal to those of 
$\mathcal{V}_{\rm phys}$ and have vanishing contribution to the matrix
elements of $T$ due to the quartet character of the states in 
$\mathcal{V}_0$. Hence as matrix elements of $T$ survive only such vectors
which are created by the particle representatives of the $r=2,0; K=L=T$
components of $\phi^{\mu\nu}$ and tree contributions of $\Sigma^{\mu\nu}$.\\

The disappearance of all contributions of the field $\Sigma^{\mu\nu}$ in the 
matrix elements of $S$- and $T$-operators beginning in one-loop has 
the effect that 
standard considerations of the optical theorem can also be extended
to the tree approximation, however with some loss of positivity there
since we are not allowed to put $c_1=c_2=0$.
This loss prevails in all contributions brought in by the tree approximation.
Apart from those the theory is reduced to EH and there the optical theorem is
valid as just explained on the quartets built with $\phi,c,\bar{c},b$. 
This also means that there is no problem
of fitting within the non-linearity of the optical theorem in 
terms of $T$. Even when using cutting rules for an explicit check
of unitarity one finds that, whenever internal $\Sigma \Sigma$ lines are cut,
those contributions contain still loops multiplied by tree's and then the
loop contributions vanish. The surviving pure tree contributions
violate unitarity.
\end{appendix}

\bibliographystyle{alpha}
\bibliography{operator_weyl}

\end{document}